\documentclass[referee]{aa} 
\usepackage{graphicx}
\begin{document}
   \title{Comments on a faint old stellar system at 150 kpc}


   \author{Hideyuki Kamaya
          \inst{1}
          }

   \offprints{H. Kamaya}

   \institute{Department of Earth and Ocean Sciences, School of Applied Sciences, National Defense Academy of Japan, Yokosuka 239-8686, Japan \\
              \email{kamaya@nda.ac.jp}
             \thanks{ Department of Astronomy, School of Science, Kyoto
   University, Sakyo-Ku, Kyoto 606-8502, Japan before 2007 April}
             }

   \date{Received February 28, 2007; accepted }

 
  \abstract
   {Recent study of SDSS J1257+3419 has revealed that this stellar system
is either a faint and small dwarf galaxy or a faint and widely extended
globular cluster. }
   {In this short note, the author suggests 
that this system is one of the smallest dwarf spheroidals (dSphs) in the
Milky Way. }
   {We re-examine some observational quantities of this object and 
check whether it can be bound system.}
   { As a result, we find the mass of SDSS J1257+3419 is the lowest of dSphs in the Milky Way, and its mass density is typical of dSphs. 
Important is that the tidal radius
of SDSS J1257+3419 is much larger than its half-light radius. That is, this very small dSph can be bound by its own gravity.}


   \maketitle
%

\section{Introduction}

Dwarf spheroidals (dSphs) are low-luminosity,
low-surface-brightness dwarf elliptical galaxies
(e.g. Irwin \& Hatzidimitriou 1995).
Their absolute visual magnitude is less than $-14$.
Their shape is near spherical. Furthermore, they do not have
 a distinct nucleus.
They were often thought in the past to be merely
large, low-density globular clusters. Recent studies have shown that
dSphs have a more complex stellar population than what is found in
globular clusters (e.g. Unavane, Wyse, \& Gilmore 1996).

Significantly, dSphs show evidence of star
formation over extended periods, even if they show no sign of
current or recent star formation and have no detectable interstellar
matter(e.g. van den Bergh 1999). 
The stellar population of dSphs consists of two
basic components: an old metal-poor population, similar to that of
globular clusters, and an intermediate-age population, whose ages range
from one to 10 billion years. The first evidence indicates
that dSphs have experienced a kind of starburst
and a galactic wind (e.g. Gallart et al. 1999).

According to the classical papers by Saito (1979a, b)
and the recent paper by de Rijcke et al. (2005), dSphs
in the Milky Way are indeed stellar systems that experienced the galactic
wind era (Dekel and Silk 1986). This is because their mass density is
relatively low although their mass is similar to globular
clusters. This indicates that the ejected interstellar medium from the
dSphs is widely distributed in the Milky Way halo (e.g. Mayer et al. 2006). 
For example, the halo medium is polluted by metal supplied by
 the galactic winds of dSphs. Then, it is
very important to search for dSphs around the Milky
Way. Especially, we need to know how many very small dSphs there are
(e.g. Read, Pontzen, \& Viel 2006).

Interestingly, recent analysis (Sakamoto \& Hasegawa 2006)
has shown there is a very small object at 150 kpc from the Sun: 
SDSS J1257+3419. This system can be either a
faint and small dwarf galaxy like dSph or a faint and widely extended globular
cluster according to the stellar properties of the system.
According to Saito (1979a), if we want to know whether SDSS J1257+3419 is a dSph, 
we should study SDSS J1257+3419 in the context of the mass and mass-density relation.
Thus, following Saito (1979a), 
we try to re-examine what the stellar system SDSS J1257+3419 is.

\section{Mass and mass density}

First of all, we estimate the mass of SDSS J1257+3419, assuming
this is a dwarf spheroidal. The
absolute magnitude of the system is $-4.8$ in V-band. Although there is
significant uncertainty in the mass-to-light ratio, $M/L$, of dwarf spheroidals,
we adopt 1.7 in the solar unit, 
because we want to compare our estimation to the results of
the original papers of Saito (cf. Kroupa 1997). 
Since the absolute magnitude of the Sun is
4.83 in V-band,
we estimate the mass of SDSS J1257+3419 to be 
$1.2 \times 10^4$ solar mass.

The mass density of SDSS J1257+3419 is also roughly estimated. 
Its half-light radius, $r_{1/2}$, is 38 pc with a large
uncertainty. Recognizing the uncertainty, I adopt this
quantity as a radius because this would not change conclusion of this note 
qualitatively. 
Hypothesizing that the system is spherical and the mass within $r_{1/2}$ is half of the total mass, the mass density
is roughly estimated as $2.6 \times 10^{-2}$ solar mass per cubic pc. This
physical dimension of density is then adopted so as to compare our study to Saito's classical discussions.

\begin{table}[t]
  \begin{tabular}{|c|c|c|c|c|}

\hline

  dSph &  distance (kpc) & Mv & $r_{1/2}$ (kpc) & M/L \\

\hline

SDSSJ1257  & 150   &    -4.8 &  0.038 & -- \\

Bootes         &  60    &   -5.7  &  0.22 & 150 \\
UrsaMinor      & 66     &   -8.9   &  0.15 & 95 \\
Sculptor       &  79    &  -11.1   &  0.094 & 11 \\
Draco          &  82    &   -8.8  &  0.12 & 243 \\
Sextans        &  86    &   -9.5  &  0.29 & 107 \\
Carina         & 101    &   -9.3  &  0.14 & 59 \\
Fornax         & 138    &  -13.2   &  0.34 & 7 \\
LeoII          & 205    &   -9.6   &  0.12 & 23 \\
CanesVenatici  & 220    &   -7.9   &  0.55 & -- \\
LeoI           & 250    &  -11.9   &  0.12 & 1.0 \\

\hline

46 Tucanae & 4.6 & -9.4 & 0.0073 & 2.0 \\

\hline
  \end{tabular}
\caption{dSph sample: data are taken from Irwin \& Hatzidimitiou (1995), except $r_{1/2}$ of Bootes (Zucker et al. 2006), $r_{1/2}$ of Canes Venatici (Belokurov et al. 2006), and M/L of Bootes (Munoz et al. 2006). Data on 46 Tucanae, a globular cluster, are taken from Pryor \& Meylan (1993).}\label{table1}
\end{table}

We adopt the well-known dSph sample listed in Table 1, including
SDSS J1257+3419. These are selected because their $r_{1/2}$
are measured well. Then, well-known dSphs Ursa Major
and Sagittarius are omitted from this list. Obviously, SDSSJ1257 denotes
SDSS J1257+3419.
Here, we call the other ten in the sample classical dSphs.

\section{Discussion and conclusion}

We first summarize  the
classical paper of Saito (1979a,b). After adopting seven dwarf spheroidals,
he found the massfs range is about $10^5 \sim 10^8$ solar mass by assuming $M/L = 1.7$. The
range of the mass density is about $10^{-3} \sim 10^{-1}$ solar mass per
cubic pc. Then, he found that the mass density of dwarf spheroidals
is much lower than that of the globular clusters. 

Both quantities of
SDSS J1257+3419 are about $10^4$ solar mass and $5.3\times 10^{-2}$
solar mass per cubic pc. Thus, we find the mass density of SDSS
J1257+3419 is not significantly different from that of classical
dSphs, although its mass is the lowest. Furthermore, since the typical mass density of globular clusters is about 1000 solar mass per cubic pc, we find its mass density is much lower than that of globular clusters.

Since SDSS J1257+3419 is very less massive, it may be a dissolving system
 due to the tidal force from the Milky Way. 
Is SDSS J1257+3419 bound by its own gravity?
Fortunately, according to a simple order of magnitude estimation,
the tidal radius of SDSS J1257+3419 is 230 pc. 
Then, we find that its $r_{1/2}$ is much smaller 
than the tidal radius. This means SDSS J1257+3419 can sustain
its structure by its own gravity
(e.g. Gonz$\rm \acute{a}$lez-Garc$\rm \acute{i}$a, Aguerri, \& Balcells 2005).
Thus, this author concludes that 
SDSS J1257+3419 can be one of the smallest dSphs in the Milky Way.

The tidal effect can be important for the formation of dSph. Mayer et al. (2001a,b)
 show that the strong tidal field of the Milky Way determines the severe mass loss in their halos and disks and induces bar and bending instabilities that transform low surface-brightness dwarfs into dSphs. If so, the size of dSphs is always smaller than the tidal radius. In the same way, SDSS J1257+3419 can be a remnant of a low surface-brightness dwarf suffering the tidal stripping.

We check the smallness of SDSS J1257+3419 from Fig. 1.
Obviously, both $r_{1/2}$ and the absolute magnitude of SDSS J1257+3419 are the smallest of sample dSphs.
For comparison, 47 Tucanae globular cluster is also over-plotted. 
The combined light from all cluster stars usually leads to a typical globular cluster with an absolute magnitude with a range of $-5$ to $-10$. 
Here, we note the absolute magnitude of SDSS J1257+3419 is smaller than for the large globular clusters like 47 Tucanae, while its mass density is much lower than that of the typical clusters thanks to the largeness of $r_{1/2}$.

\begin{figure}[t]
  \scalebox{0.70}{
      \includegraphics{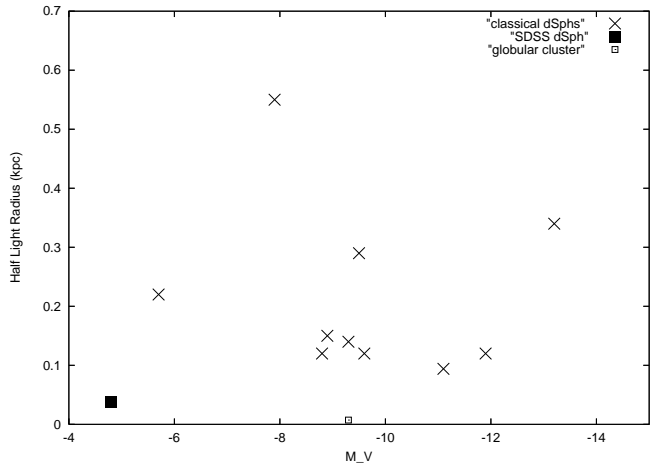}
          \label{fig:dsph}
            }
\caption{This figure indicates SDSS J1257 is one of the smallest dwarf
            spheroidal galaxies. Crosses are the classical dwarf spheroidals,
            while the square is SDSS J1257. Small open square is 47 Tucanae (a globular cluster). Vertical axis is half light radius,
            and horizontal axis is absolute magnitude in V-band.}
\begin{tabular}{p{7.0cm}}
\\ \hline
\end{tabular}
\end{figure}

The most uncertain point in our discussion is $M/L$. 
The mass-to-light ratio of dSphs is usually 
higher than that of globular clusters as shown in Table 1.
However, as is well known, $M/L$ of dSphs is significantly
debating (e.g. Kroupa 1997). At the worst case, the uncertainty of $M/L$ is about a factor of two. In the case of SDSS J1257+3419,
if its $M/L$ is 100, 
the mass density can reach at $1.5$ solar mass per cubic pc.
Thus, when we want to know the realistic mass density of the current object, more  precise observation is necessary.
Fortunately, the qualitative sequence is not varied as summarized in Fig. 1, which suggests SDSS J1257+3419 is one of the smallest dSphs, because quantities of both the axes are not affected by the uncertainty of $M/L$.
And, the mass density with higher $M/L \sim 100$ is still much lower than that of globular clusters. 
Thus, our conclusion is not altered.

\section{Summary}
The recent analysis of SDSS J1257+3419 has suggested
 that this stellar system
is either a faint and small dwarf galaxy or a faint and widely extended
globular cluster. 
According to our study, the mass of SDSS J1257+3419 is the lowest of the sample dSphs, and the mass density is very similar to that of classical dSphs.
As a result, the author insists this system is one of the smallest dSphs in the Milky Way.

\begin{acknowledgements}
I have appreciated the productive comments of the referee, which improves the paper very much.
\end{acknowledgements}

\end{document}